\documentclass[conference]{IEEEtran}  
\usepackage{ragged2e}
\usepackage{floatrow}
\usepackage{multirow}
\usepackage{amsthm}
\usepackage[utf8]{inputenc}
\usepackage[english]{babel}
\theoremstyle{remark}
\newtheorem{theorem}{Theorem}

\newtheorem{corollary}{Corollary}
\usepackage{multicol}
\usepackage{epsfig}
\usepackage{epstopdf}
\usepackage{float}
\usepackage{perpage}
\MakeSorted{figure}
\MakeSorted{table}
\usepackage{array}
\usepackage{graphicx}
\usepackage{stfloats}
\usepackage{amsfonts}
\usepackage{amssymb}
\usepackage{soul}

\usepackage{enumitem}							
\usepackage[english]{babel}
\usepackage[utf8]{inputenc}
\usepackage[linesnumbered,ruled,vlined]{algorithm2e}
\usepackage{upgreek}
\usepackage{bm} 
\usepackage{hyperref}
\usepackage{float}
\floatstyle{plaintop}
\restylefloat{table}
\hypersetup{linktocpage} 
\hypersetup{
colorlinks,
citecolor=black,
filecolor=black,
linkcolor=black,
urlcolor=black
}
\def\BibTeX{{\rm B\kern-.05em{\sc i\kern-.025em b}\kern-.08em
    T\kern-.1667em\lower.7ex\hbox{E}\kern-.125emX}}
\usepackage{pifont}
\usepackage{makecell}


\usepackage{cite}
\DeclareMathAlphabet\mathbfcal{OMS}{cmsy}{b}{n}
\ifCLASSINFOpdf
\else
\fi
\usepackage[cmex10]{amsmath}

\usepackage{array}
\usepackage{fixltx2e}

\usepackage[linesnumbered,ruled,vlined]{algorithm2e}

\usepackage[font=footnotesize,labelfont=bf, figurename=Fig.]{caption} 
\usepackage{subcaption}
\usepackage{fix-cm}
\usepackage{comment}

\usepackage{mathtools}
\usepackage{pifont}

\newcommand{\Amat}{\mathbf{A}}
\newcommand{\Bmat}{\mathbf{B}}
\newcommand{\Cmat}{\mathbf{C}}
\newcommand{\Dmat}{\mathbf{D}}

\newcommand{\Imat}{\mathbf{I}}

\newcommand{\Pmat}{\mathbf{P}}
\newcommand{\Qmat}{\mathbf{Q}}
\newcommand{\Rmat}{\mathbf{R}}
\newcommand{\Smat}{\mathbf{S}}

\newcommand{\Wmat}{\mathbf{W}}

\newcommand{\av}{\mathbf{a}}
\newcommand{\bv}{\mathbf{b}}

\newcommand{\fv}{\mathbf{f}}

\newcommand{\hv}{\mathbf{h}}
\newcommand{\nv}{\mathbf{n}}

\newcommand{\yv}{\mathbf{y}}
\newcommand{\zv}{\mathbf{z}}

\newcommand{\E}{\mathbb{E}}

\begin{document}

\title{{LSFD for Rician-Faded Cell-Free  mMIMO Systems With Channel Aging and Hardware Impairments}}
\author{\\[-20pt]
	\IEEEauthorblockN{Anish Chattopadhyay, Venkatesh Tentu, Dheeraj Naidu Amudala and Rohit Budhiraja}
	\IEEEauthorblockA{Department of Electrical Engineering, Indian Institute of Technology Kanpur  
		\\\{anishchat20, tentu, dheeraja, rohitbr\}@iitk.ac.in\\[-15pt] 
	}
}
\maketitle
\vspace{-1.99cm}
\begin{abstract}
We study the impact of channel aging on the uplink of a cell-free massive multiple-input multiple-output system with hardware impairments. We consider a dynamic analog-to-digital converter architecture at the access points (APs), and  low-resolution digital-to-analog converters at the user equipments (UEs). We derive a closed-form spectral efficiency expression~by considering i) practical spatially-correlated Rician channels; ii) hardware impairments  at the APs and the UEs; iii) channel~aging; and iv) large-scale fading decoding (LSFD). We show that LSFD can effectively mitigate the detrimental effects of i) channel aging for both low and high UE velocities; and ii) inter-user interference for low-velocity UEs but not for high-velocity UEs.
\end{abstract}
\vspace{-.25cm}
\section{Introduction}
Cell-free (CF) massive multiple-input multiple output (mMIMO) is being investigated as a key technology
for beyond fifth-generation wireless systems. This is because it can provide a uniformly high spectral efficiency (SE) in its entire coverage area~\cite{HQNgo2017}. A CF mMIMO system deploys a large number of access points (APs)~\cite{hu2019_tcom}, and avoids inter-cell interference via cooperation between APs, which are inter-connected using a  central processing unit (CPU)~\cite{HQNgo2017,hu2019_tcom}. 

A CF mMIMO system, where APs and user equipments (UEs)  are designed  using high-quality radio frequency (RF) transceivers and high-resolution analog-to-digital converters (ADCs)/digital-to-analog converters (DACs), have a high implementation cost and energy consumption~\cite{hu2019_tcom,zhang_21_ADC_DAC,zhang_20_mixed_ADC_DAC}.
A CF mMIMO deployment is attractive when APs and UEs are designed using inexpensive RF transceivers and low-resolution ADC/DACs \cite{HQNgo2017,hu2019_tcom}.  To reduce the SE loss due to low-resolution ADC/DACs, the authors in~\cite{zhang_21_ADC_DAC,zhang_20_mixed_ADC_DAC} considered a mixed-ADC architecture, wherein one fraction of AP antennas have a high-resolution ADCs, while the other has low-resolution. Zhang \textit{et. al} in \cite{zhang_21_ADC_DAC} and \cite{zhang_20_mixed_ADC_DAC} investigated the SE of a CF mMIMO system with mixed-ADC architecture at the APs by considering {spatially-uncorrelated Rayleigh and Rician-fading channels respectively, and showed its superiority over the low-resolution counterpart.}
The current work generalizes the mixed-ADC architecture in  \cite{zhang_20_mixed_ADC_DAC,zhang_21_ADC_DAC}, by considering a dynamic ADC architecture, wherein the  resolution of ADC connected to each antenna can be varied from $1$ to maximum $b$ bits. This is in unlike~\cite{zhang_20_mixed_ADC_DAC,zhang_21_ADC_DAC}, wherein each AP antenna can either have a low or a high-resolution. 

To further reduce the cost and energy consumption of CF mMIMO systems, the authors in \cite{Masoumi_CF_HW_20,CF_UAV_22} assumed that the APs and UEs additionally employ low-cost RF transceivers. The low-cost RF transceivers, however, experience power amplifier non-linearities and phase noise, which degrade the benefits accrued from CF mMIMO technology~\cite{Masoumi_CF_HW_20,CF_UAV_22}.
  Masoumi \textit{et. al} in~\cite{Masoumi_CF_HW_20}  derived the uplink SE of a hardware-impaired uncorrelated Rayleigh-faded CF mMIMO system with limited-capacity fronthaul link between each AP and CPU. 
  Tentu \textit{et. al} in \cite{CF_UAV_22} investigated the SE of a spatially-correlated Rician-faded CF mMIMO system with low cost RF transceiver. 

The aforementioned works analyzed the impact of low and mixed resolution ADCs and low-cost RF chains for Rayleigh-faded channels, except \cite{zhang_20_mixed_ADC_DAC,CF_UAV_22} which considered Rician-faded channels.  Rayleigh fading may not accurately characterize the propagation environment, as the UE-AP channel may contain line-of-sight (LoS) path due to its macro-diversity~\cite{Jin_CF_EXP_CORR_20, Emil2019_rician_phase}. 
Such a channel has a Rician probability distribution function (pdf). The phase of LoS path, which accounts for the phase-shift due to UEs mobility, is modeled~as a uniformly distributed random variable~\cite{Emil2019_rician_phase}. 
The authors in \cite{Jin_CF_EXP_CORR_20,CF_UAV_22} did not consider phase-shifts in the LoS component while modeling the Rician channel. 
Jin \textit{et. al} in~\cite{Jin_CF_EXP_CORR_20} derived a closed-form SE expression for spatially-correlated Rician-faded CF mMIMO system by assuming  perfect knowledge of phase-shifts.  In practice,  phase-shifts are hard to estimate~\cite{Emil2019_rician_phase}.  The authors in~\cite{Emil2019_rician_phase} investigated the uplink and downlink~SE of~a uncorrelated Rician-faded CF mMIMO system by assuming unknown phase-shift, \textit{but with ideal hardware at the APs and~UEs.} 


The UEs to AP channels are further subjected to time variations due to UE mobility, a phenomenon known as channel aging~\cite{chopra_AGING_21,Zheng_aging_correlated_21}. 
 Chopra \textit{et. al} in \cite{chopra_AGING_21} showed that the impact of channel aging is higher for a CF mMIMO than its counterpart.  
 Zheng \textit{et. al} in \cite{Zheng_aging_correlated_21} analyzed the effect of channel aging in a CF mMIMO system for correlated Rayleigh fading channels. 
 The authors in~\cite{chopra_AGING_21,Zheng_aging_correlated_21}, however, assumed cellular/CF mMIMO systems with ideal hardware. The current work analyzes the SE of a spatially-correlated Rician-faded CF mMIMO system with random LoS phase-shifts, low-cost RF chains, and dynamic-ADC/DAC architecture. 

{It is practically  difficult, due to the limited coherence block,  to assign orthogonal pilots to each user. This pilot sharing across users leads to pilot contamination. A two-layer  large-scale fading decoding (LSFD) is proposed to combat it~\cite{van_19_mcell_LSFD,Ozlem_CFwpt_21}. 
 Trinh \textit{et. al} in \cite{van_19_mcell_LSFD} showed that in Rayleigh-faded cellular mMIMO systems LSFD significantly reduces the impact of pilot contamination, which conserably improves the SE. Ozlem \textit{et. al} in \cite{Ozlem_CFwpt_21} maximized the minimum SE in a wireless-powered Rician-faded CF mMIMO system with LSFD. The authors in~\cite{van_19_mcell_LSFD,Ozlem_CFwpt_21}, however assumed ideal hardware.}
\textit{The existing CF mMIMO works, to the best of our knowledge have not even investigated the performance of a hardware-impaired CF mMIMO system with spatially-correlated Rician-fading channel and aging.}
We next summarize the \textbf{main contributions} of the current work, which address these gaps:\newline 
$\bullet$ We consider a CF mMIMO system with spatially-correlated Rician-faded channels with aging, and investigate the impact of dynamic-ADC architecture and low-cost hardware-impaired RF chains. We also consider a two-layer LSFD and derive a closed-form SE expression by addressing the derivation difficulties caused by the combined modelling of channel aging, LSFD, RF impairments, dynamic ADC/DAC architecture and spatially-correlated Rician channels.\newline   
$\bullet$ We show that LSFD, even in presence of hardware impairments and channel aging,  provides high SE gain over single layer decoding. {Further, LSFD  effectively mitigates the degradation due to channel aging for both low and high UE
velocities. It can, however, mitigate inter-user interference (IUI) only for low-velocity UEs,  and not for high-velocity~UEs.}
\section{System Model}
We consider the uplink of a CF mMIMO system with $M$ multi-antenna APs, each equipped with $N$ antennas,  and $K$ single-antenna UEs. We assume, similar to \cite{Ozlem_CFwpt_21, hu2019_tcom}, that the APs are randomly distributed over a large geographical area, and are connected to a CPU via high-speed fronthaul links. To reduce the system hardware cost and power consumption, both APs and UEs are equipped with cost-effective RF chains. Further, the APs have a dynamic-resolution ADC architecture, wherein each AP antenna can be connected to a different resolution ADC. The  UEs are designed using low-resolution DACs. This is unlike  \cite{hu2019_tcom,zhang_20_mixed_ADC_DAC,zhang_21_ADC_DAC}
, which model all the ADCs as either low- or mixed-resolution. 

We also assume that the CF mMIMO system, due to UEs mobility, observes channel aging. To investigate its impact,  we consider a resource block, each of length $\tau_c$ time instants {\cite{Zheng_aging_correlated_21}}.  The channel remains constant for a time instant, and varies across time instants in a correlated manner. This temporal correlation is modeled later using Jake's model~\cite{chopra_AGING_21}.  We also assume that each resource block is divided into an uplink training and data transmission interval of lengths $\tau_p$ and $(\tau_c-\tau_p)$ time instants, respectively.  We next discuss the UE-AP uplink channel, and then the uplink training and data transmission protocols.


\underline{\textbf{Channel model:}} The  channel from the $k$th UE to the $m$th AP at the $\lambda$th time instant is denoted as  $\hv_{m,k}[\lambda] \in \mathbb{C}^{N\times 1}$. Due to dense AP deployment, $\hv_{m,k}[\lambda]$ contains  both LoS and non-LoS paths, and, therefore, has a  Rician pdf \cite{Emil2019_rician_phase} i.e., 
\begin{align}
    \hv_{mk}[\lambda] = \Bar{\hv}_{mk} e^{j\phi_{mk}^{\lambda}} + \Rmat_{mk}^{1/2}\tilde{\hv}_{mk}[\lambda].  
\end{align} 
Here $ \Bar{\hv}_{mk} \!=\! \sqrt{\frac{ K_{mk}\beta_{mk}}{K_{mk}+1}} \Breve{\hv}_{mk}$ and $\Rmat_{mk} \!= \!\sqrt{\frac{\beta_{mk}}{K_{mk}+1}}\tilde{\Rmat}_{mk}$.
The term $K_{mk}$ is the Rician factor, and $\beta_{mk}$ is  the large-scale fading coefficient. 
 The LoS term $ \Breve{\hv}_{mk}$ is modeled  as $ \Breve{\hv}_{mk}=[1, e^{j\psi_{mk}}, \cdots, e^{j(N-1)\psi_{mk}}]$, where $\psi_{mk}$ is the angle of arrival between the $m$th AP and the $k$th UE. The term, $\tilde{\hv}_{mk}[\lambda] $, with pdf $\mathcal{CN}(\mathbf{0},\mathbf{I}_{N})$, models the small-scale fading, and $\tilde{\Rmat}_{mk}$ characterizes the spatial correlation matrix of the NLoS path. The phase-shift  $\phi_{mk}^{\lambda}$ at the $\lambda$th instant is uniformly distributed between $[-\pi, \pi]$. 
 	
The long-term channel statistics i.e., $\bar{\hv}_{mk}$, $\beta_{mk}$ and $\Rmat_{mk}$, remain constant over a resource block and, similar to the existing literature~\cite{Emil2019_rician_phase,Ozlem_CFwpt_21}, are perfectly known at the AP. The LoS phase-shift $\phi_{mk}^{\lambda}$, similar to small-scale fading component, varies at each time instant \cite{Emil2019_rician_phase,Ozlem_CFwpt_21}. Most of the  Rican CF works e.g.,~\cite{Jin_CF_EXP_CORR_20,CF_UAV_22}, assume a LoS path with static phase, and thus ignored its effect. A slight change in UEs position and hardware impairments can radically change the phase. It is crucial to consider this phase-shift to realistically model the Rician channel~\cite{Emil2019_rician_phase,Ozlem_CFwpt_21}.  
 We, similar to~\cite{Emil2019_rician_phase,Ozlem_CFwpt_21}, model the channel by assuming that the LoS phase varies as frequently as small-scale fading, and consequently the AP is unaware of it.  As an AP does not have prior knowledge of $\phi_{mk}^{\lambda}$, it estimates the channel $\hv_{ak}$, without its knowledge.

The UEs mobility causes the AP-UE channel in a {resource block to vary across the time instants, which
causes channel aging~\cite{Zheng_aging_correlated_21}. We model~the channel $\hv_{mk}[n]$ at the $n$th time instant,  as a combination of its initial channel $\hv_{mk}[0]$,  and innovation component as follows \cite{Zheng_aging_correlated_21}:\vspace{-3pt}
\begin{align} \setcounter{equation}{1}
  \!\! \!\! \hv_{mk}[n] \!=\!\rho_{k}[n]\hv_{mk}[0] \!+ \!\sqrt{\!1\!-\!\rho_{k}^{2}[n]} \Big( \Bar{\hv}_{mk}e^{j\phi_{mk}^{n}} \!+ \fv_{mk}[n] \Big).\!\! \label{eq_channel_h[n]}
\end{align}
Here $\rho_{k}[n]$ is the temporal correlation coefficient, which based on the Jake's model~\cite{Zheng_aging_correlated_21}, is given as $\rho_{k}[n]= J_{0}(2\pi f_{d,k}T_s n)$.  The term $J_{0}(\cdot)$ is the zeroth-order Bessel function of the first kind, and $T_s$ is the sampling time. The term  $f_{d,k}=(v_{k}f_{c})/c$ is the Doppler spread, with $v_{k}$, $f_{c}$ and $c$ being the user velocity, carrier frequency and the velocity of light, respectively. The innovation component $\fv_{mk}[n]$ is independent of the channel ${\hv}_{mk}[0]$, and has a pdf $\mathcal{CN}(\mathbf{0},\Rmat_{mk})$~\cite{Zheng_aging_correlated_21}.\newline
\textbf{Uplink training:}
Recall that the uplink training phase consists of $\tau_p$ time instants. The $k$th UE, similar to~\cite{Zheng_aging_correlated_21}, transmits its pilot signal $\sqrt{\tilde{p}_{k}}\phi_{k}[t_k]$ at the time instant $t_k\subset \{1, \dots, \tau_p\}$. Here  $|\phi_{k}[t_k]|^2=1$. 
We assume that the uplink training duration $\tau_p< K$. The number of UEs transmitting pilot at a particular time instant is more than one, which causes pilot contamination~\cite{Emil2019_rician_phase}. The set of UEs that transmit pilots at the time instant $t_k$ is denoted as  $\mathcal{P}_k$.   The $k$th UE feeds its pilot signal to the low-resolution DAC, which distorts it.  This distortion is commonly analyzed using Bussgang model~\cite{demir_bussgang_21}. The distorted DAC output, based on the Bussgang model, is:\vspace{-3pt}
\begin{align}
s_{\text{DAC},k}[t_k]=Q(\sqrt{\tilde{p}_{k}}\phi_{k}[t_k])=\alpha_{d,k}\sqrt{\tilde{p}_{k}}\phi_{k}[t_k]+\upsilon_{\text{DAC},k}[t_k].\notag
\end{align}
Here $\alpha_{d,k} \!= \! 1\!-\!\rho_{d,k}$ is the DAC distortion factor and $\upsilon_{\text{DAC},k}[t_k]$ is the zero-mean DAC quantization noise, which is uncorrelated with the input pilot signal \cite{Emil2019_rician_phase}. Its variance is given as $\rho_{d,k}\alpha_{d,k}\tilde{p}_{k}\mathbb{E}\big(|\phi_{k}[t_k]|^2\big)$. The $k$th UE DAC output signal $s_{\text{DAC},k}[t_k]$ is then fed to its low-cost hardware-impaired RF chain whose output, based on the error vector magnitude (EVM) model, adds a distortion term to the transmit signal~\cite{mMIMObook2017}.  The effective uplink pilot signal is, therefore, given as $s_{\text{RF},k}[t_k] =  s_{\text{DAC},k}[t_k] +\eta^{\text{UE}}_{t,k}[t_k]$. Here $\eta^{\text{UE}}_{t,k}[t_k]$ is the distortion term, which is independent of the input signal, and has a pdf $ \mathcal{CN}\big(0,\kappa_{t,k}^{2}\big(\mathbb{E}\{s_{\text{DAC},k}[t_k]s_{\text{DAC},k}^{H}[t_k]\}\big)\big)$~\cite{mMIMObook2017}. The term $\kappa_{t,k}$ models the UE transmit EVM~\cite{mMIMObook2017}.
\begin{figure*}[t!]
\setcounter{equation}{5}
\vspace{3pt}
\begin{align}
\!\!\!\! \yv_{\!\text{ADC},m}^{p}[t_k]\! =\!\sum_{i\in \mathcal{P}_k}\!\!\Amat_{m}\!\left(\! \rho_{i}[t_k^{\lambda}]\hv_{mi}[\lambda]\! +\!\Bar{\rho_{i}}[t_k^{\lambda}]\breve{\fv}_{mi}[t_k] \right)\! \Big(\!{\alpha_{d,i}\sqrt{ \!\tilde{p}_{i}}} \! +\!n_{\text{DAC},i}[t_k]\!
 + \!\xi_{\text{RF},i}[t_k]\! \Big)\! +\!\Amat_{m} \big(\!\boldsymbol{\eta}_{\text{RF},m}[t_k] \! +\!\zv_{m}[t_k]\big)\!+\!\boldsymbol{n}_{\text{ADC}}^{m}[t_k].\!\! \label{eq:y_m[t_k]} 
\end{align}
\hrule
\vspace{-15pt}
\end{figure*}

The pilot signals received  at the antennas  of the $m$th AP at the  time instant $t_k$ is\vspace{+1pt} 
\begin{align} \setcounter{equation}{2}
\yv_{m}^p[t_k]=\sum\limits_{k\in\mathcal{P}_{k}}\hv_{mk}[t_k]s_{\text{RF},k}[t_k].\label{eq_AP_rx_signal_without_HW}
\end{align}
 To reduce the system cost, APs are designed using low-cost hardware-impaired RF chains and a dynamic-ADC architecture, which enables us to vary the resolution of each ADC from $1$ to maximum of $b$ bits. The $m$th AP feeds the pilots signal received at its antenna to the low-cost RF chains, whose distorted output, based on the EVM model, is {\cite{CF_UAV_22}:\vspace{-3pt}  
\begin{align}
\yv_{\text{RF},m}^p[t_k] =  \yv_{m}^p[t_k]  +\boldsymbol{\eta}_{\text{RF},m}[t_k]+\zv_{m}[t_k].\label{eq_RX_signal_AP_RF_HW1}
\end{align}
The term $\boldsymbol{\eta}_{\text{RF},m}[t_k]$, with pdf $\mathcal{CN}(\boldsymbol{0},\kappa_{r,m}^{2}\Wmat_{m}[t_k])$, models the hardware distortion due to the low-cost RF chains. The term $\kappa_{r,m}^{2}$ is the AP receiver EVM and $\Wmat_{m}[t_k] = \text{diag}\big(\E\big\{\yv_{m}^p[t_k](\yv_{m}^p[t_k])^{H}\big| \hv_{mk}[t_k]\big\}\big)$. The vector $\zv_{m}[t_k]$, with pdf  $\mathcal{CN}(\boldsymbol{0},\mathbf{I}_N)$, is the AWGN at the $m$th AP.  The RF chain output is then fed to the dynamic resolution ADCs, which distorts it by adding quantization noise. The distorted ADCs output, based on the Bussgang model~\cite{demir_bussgang_21}, is given as\vspace{-3pt} 
\begin{align}
&\yv_{\text{ADC},m}^{p}[t_k]= \Amat_{m}\yv_{\text{RF},m}^p[t_k]+{\nv_{q}^{m}}[t_k] \label{eq_RX_signal_AP_RF_HW_ADC}.
\end{align}
 The matrix $\Amat_{m}\!=\!\text{diag}(1\!-\!\rho_{a,1},\cdots,1\!-\!\rho_{a,N})$, where  $\rho_{a,n}$ is  the ADC distortion factor~\cite{demir_bussgang_21}. The zero-mean additive ADC quantization noise ${\nv_{q}^{m}}[t_k]$ is uncorrelated with $\yv_{\text{RF},m}^{p}[t_k]$. Its covariance is $\Bmat_{m}\text{diag}\big(\mathbb{E}\big \{\yv_{\text{RF},m}^{p}[t_k](\yv_{\text{RF},m}^{p}[t_k])^{H}\big|\hv_{mk}[t_k]\big \}\big)$ with $\Bmat_{m}\!=\!\Amat_{m}(\Imat_{N}\!-\!\Amat_{m})$~\cite{demir_bussgang_21}. The existing CF mMIMO literature~\cite{hu2019_tcom,zhang_20_mixed_ADC_DAC,zhang_21_ADC_DAC}, has not investigated the dynamic resolution ADC architecture with different diagonal elements of $\Amat_m$ i.e., $\rho_{a,p} \neq \rho_{a,q}$ for $p \neq q$. 
 It significantly complicates the SE analysis and derivation, when  compared with \cite{hu2019_tcom,zhang_20_mixed_ADC_DAC,zhang_21_ADC_DAC}, 
which considers either low- or mixed-ADC architecture at the APs. 
Our architecture is generic, and reduces to its low-resolution and mixed-resolution counterparts with $\Amat_m = (1-\rho_{a})\mathbf{I}_N$  and $\Amat_m= \text{blkdiag}\{(1-\rho_{a})\mathbf{I}_{\gamma},\mathbf{I}_{N-\gamma}\}$, respectively.
%


We next substitute expressions of $\yv_{m}^{p}[t_k]$ and $\yv_{\text{RF},m}^{p}[t_k]$ from \eqref{eq_AP_rx_signal_without_HW} and \eqref{eq_RX_signal_AP_RF_HW1} in \eqref{eq_RX_signal_AP_RF_HW_ADC}, and re-express $\yv_{\text{ADC},m}^{p}[t_k]$ as\vspace{-3pt}  
\begin{align}
   \yv_{\text{ADC},m}^{p}[t_k]&= \!\sum_{i\in \mathcal{P}_k} \! \Amat_{m}\big(\!\hv_{mi}[t_k]\big({\alpha_{d,i}\sqrt{\tilde{p}_{i}}}+\upsilon_{\text{DAC},i}[t_k]\notag \\[-3pt]
  &+\!\xi_{\text{RF},i}[t_k]\big) +\!\boldsymbol{\eta}_{\text{RF},m}[t_k]  +\zv_{m}[t_k]\big)\!+\nv_{\text{ADC},m}[t_k]. \notag 
\end{align} 
Here $ \alpha_{d,i}\!=\!(1\!-\!\rho_{d,i})$. Recall that the channel between the two different time instants ages, and is consequently  correlated. The received signal $\yv_{\text{ADC},m}^{p}[t_k]$ can be exploited while estimating channel at any other time instant also. The channel estimate quality will, however, deteriorate with increasing time difference between the pilot transmission ($1<n<\tau_p$) and the considered channel realization ($\tau_p+1<n<\tau_c$). We, therefore, without loss of generality, estimate the channel at the time instant $\lambda = \tau_p +1$, and use these estimates to obtain channels at all other time instants $n >\lambda$. To obtain the channel at the $\lambda$th time instant, the received pilot signal $\yv_{\text{ADC},m}^{p}[t_k]$ is expressed in terms of the channel at the time instant $\lambda$, using \eqref{eq_channel_h[n]}  as given in \eqref{eq:y_m[t_k]} at the top of the page. 
Here $t_k^{\lambda} = \lambda - t_k$ and  $\Bar{\rho_{k}}[t_k^{\lambda}] = \sqrt{1-\rho_{k}[t_k^{\lambda}]}$.
 Using $\yv_{\text{ADC},m}^{p}[t_k]$, the channel $\hv_{mk}[\lambda]$  is estimated in the following theorem. The proof, due to page constraints, is provided in~\cite{CF_AGING_APPENDIX_DOC_22}.
\begin{theorem}\label{ch_estimate_theorem}
For a hardware-impaired CF mMIMO system with spatially-correlated Rician fading and phase-shifts, the linear minimum mean square error (LMMSE) estimate of $\hv_{mk}[\lambda]$ is:\vspace{-2pt}
\begin{align} \setcounter{equation}{6}
    \hat{\hv}_{mk}[\lambda]=\sqrt{\tilde{p}_{k}}\rho_{k}[t_k^{\lambda}]\bar{\Rmat}_{mk}\Amat_{m}\boldsymbol{\Psi}_{mk} \yv_{\text{ADC},m}^{p}[t_k]\;, \text{ where}  \label{eq: LMMSE}
\end{align}
$\boldsymbol{\Psi}_{\!mk} \!=  \!\Big(\!\sum\limits_{i\in \mathcal{P}_k}\!\! \alpha_{d,i}(1\!+\! \kappa^2_{t,i})\tilde{p}_{i}\Big(\Amat_{m}\mathbf{\Bar{R}}_{mk}\Amat_{m}^H \!
    +\!\big(\mathbf{B}_{a}^{m}\!+\! \kappa^{2}_{r,m}\Amat_{m}\big)\!$   \vspace{-2pt}
    $\;\times \text{diag}\big(\bar{\Rmat}_{mi}\big) \Big) \!+\! \sigma^{2}\! \Amat_{m}\! \Big)^{\!-1}\!$ and  $\mathbf{\bar{R}}_{mk}=\big( \bar{\hv}_{mk}\Bar{\hv}_{mk}^{H}+\Rmat_{mk}\big)$. 
\end{theorem}
\begin{figure*}[t]
\setcounter{equation}{10}
{\begin{align}\label{eq_signal_at_CPU}
  &  \Hat{s}_{k}[n]=\underbrace{ \sum_{m=1}^{M}a_{mk}^{*}[n]\alpha_{d,k}\sqrt{p_{k}}\Hat{\hv}_{mk}[\lambda]\Amat_{m}\hv_{mk}[n]s_{k}[n]}_{\text{Desired Signal, }\text{DS}_{k,n}} + \sum_{i\neq k}^{K}\underbrace{\sum_{m=1}^{M}a_{mk}^{*}[n]\alpha_{d,i}\sqrt{p_{i}} \Hat{\hv}_{mk}^{H}[\lambda]\Amat_{m}\hv_{mi}[n]s_{i}[n]}_{\text{Inter-user interference, } \text{IUI}_{ki,n}} \nonumber\\[-3pt]
    &+\underbrace{\sum_{m=1}^{M}a_{mk}^{*}[n]\Hat{\hv}_{mk}^{H}[\lambda]\Amat_{m}\Big(\sum_{i=1}^{K}\hv_{mi}[n]\upsilon_{\text{DAC},i}[n]\Big)}_{\text{UE DAC impairment, } \text{DAC}_{k,n}} +\underbrace{\sum_{m=1}^{M}a_{mk}^{*}[n]\Hat{\hv}_{mk}^{H}[\lambda]\Amat_{m}\Big(\sum_{i=1}^{K}\hv_{mi}[n]\xi_{\text{RF},i}[n]\Big)}_{\text{UE RF impairment, } \text{TRF}_{k,n}} \nonumber\\[-3pt]
    &+\underbrace{\sum_{m=1}^{M}a_{mk}^{*}[n]\Hat{\hv}_{mk}^{H}[\lambda]\Amat_{m}\boldsymbol{\eta}_{\text{RF},m}[n]}_{\text{AP RF impairment, } \text{RRF}_{k,n}}  +\underbrace{\sum_{m=1}^{M}a_{mk}^{*}[n]\Hat{\hv}_{mk}^{H}[\lambda]\nv_{\text{ADC},m}[n]}_{\text{AP ADC impairment, } \text{ADC}_{k,n}} +\underbrace{\sum_{m=1}^{M}a_{mk}^{*}[n]\Hat{\hv}_{mk}^{H}[\lambda]\Amat_{m}{\zv}_{m}[n]}_{\text{AWGN noise, } \text{NS}_{k,n}}.
\end{align}}
\hrule
\vspace{-5pt}
\end{figure*}
\begin{table*}[]   
 \footnotesize
 \centering
    \caption{Simulated expressions for the desired signal and interference terms.}  \vspace{-0.05in}
   \label{table_DS_IUI_HW_ADC}
  \begin{tabular}{|c|c|} 
 \hline
$\overline{\mbox{DS}}_{k,n}\!=\!\Big| \sum\limits_{m=1}^{M}\bar{a}_{mk}^{\rho *}[n]\sqrt{p_{k}}\E\big\{\hat{\hv}_{mk}^{H}[\lambda]\Amat_{m}\hv_{mk}[\lambda] \big\} \Big|^{2} \!$ &   $ \overline{\mbox{CA}}_{k,n}\!=\!\E\Big\{\Big| \sum\limits_{m=1}^{M}\bar{a}_{mk}^{\bar{\rho} *}[n]\sqrt{p_{k}}\hat{\hv}_{mk}^{H}[\lambda]\Amat_{m}\bar{\fv}_{mk}[\lambda]  \Big|^{2} \Big\}$ \\ 
$\overline{\mbox{IUI}}_{ki,n}\!=\!\!\E\Big\{ \Big| \sum\limits_{m=1}^{M}a_{mk}^{*}[n]\alpha_{d,i}\sqrt{p_{i}} \hat{\hv}_{mk}^{H}[\lambda]\Amat_{m}\hv_{mi}[n] \Big|^{2} \Big\}$ & $\overline{\text{RRF}}_{k,n}\!=\!\E\Big\{\Big| \sum\limits_{m=1}^{M}a_{mk}^{*}[n]\hat{\hv}_{mk}^{H}[\lambda]\Amat_{m}\boldsymbol{\eta}_{r,m}^{\text{AP}}[n] \Big|^{2} \Big\}$    \\
 $\overline{\text{TRF}}_{k,n}\!=\!\E\Big\{\Big|  \sum_{m=1}^{M}a_{mk}^{*}[n]\hat{\hv}_{mk}^{H}[\lambda]\Amat_{m}\hv_{mi}[n]\xi_{\text{RF},i}[n]  \Big|^{2} \Big\}\!$ &  $\overline{\text{ADC}}_{k,n}=\E\Big\{ \Big| \sum\limits_{m=1}^{M}a_{mk}^{*}[n]\hat{\hv}_{mk}^{H}[\lambda]\nv_{\text{ADC},m}[n] \Big|^{2} \Big\}$ \\  
 $\overline{\mbox{DAC}}_{k,n}=\E\Big\{\Big|  \sum\limits_{m=1}^{M}a_{mk}^{*}[n]\hat{\hv}_{mk}^{H}[\lambda]\Amat_{m}\hv_{mi}[n]\upsilon_{\text{DAC},i}[n]  \Big|^{2} \Big\}$ & $\overline{\text{NS}}_{k,n}\!\!= \E\Big\{ \Big|\sum\limits_{m=1}^{M}a_{mk}^{*}[n]\hat{\hv}_{mk}^{H}[\lambda]\Amat_{m}\zv_{m}[n] \Big|^{2}\Big\}$  \\ 
 \hline 
        \multicolumn{2}{|c|}{$\overline{\mbox{BU}}_{k,n}\!=\!\E\Big\{\Big| \sum\limits_{m=1}^{M} \bar{a}_{mk}^{\rho *}[n] \sqrt{p_{k}}\Big(\hat{\hv}_{mk}^{H}[\lambda]\Amat_{m}\hv_{mk}[\lambda] - \E\big\{\hat{\hv}_{mk}^{H}[\lambda]\Amat_{m}\hv_{mk}[\lambda] \big\} \Big) \Big|^{2}  \Big\}\!$, with $ \bar{a}_{mk}^{\rho *}[n] = a_{mk}^{*}[n]\alpha_{d,k}\rho_{k}[n_{\lambda}]$.} \\
 \hline
  \end{tabular}
  \vspace{-15pt}
\end{table*}
The LMMSE estimation error $\tilde{\hv}_{mk}=\hv_{mk}-\hat{\hv}_{mk}$ has zero mean, and  covariance matrix $\Cmat_{mk} = \bar{\Rmat}_{mk}- \alpha_{d,k}^2\tilde{p}_{k}\rho_{k}^2[t_k^{\lambda}]\bar{\Rmat}_{mk}\Amat_{m} \boldsymbol{\Psi}_{mk}\Amat_{m}^H \bar{\Rmat}_{mk}^H $. The estimate $\hat{\hv}_{mk}$ is uncorrelated with the error $\tilde{\hv}_{mk}$.
The channel estimator derived in \cite{Zheng_aging_correlated_21}, by assuming Rayleigh fading and ideal RF chains and ADC/DAC, cannot be used herein. This is due to the hardware-impaired RF chains and dynamic-resolution ADC/DAC architecture, we need to compute $\boldsymbol{\Psi}_{mk}$, which \cite{Zheng_aging_correlated_21} need not.
\textbf{Uplink Transmission:}
Let $s_{k}[n]$, with $\mathbb{E}\{|s_{k}[n]|^2\}=1$, be the information symbol which the $k$th UE wants to transmit at the $n$th  time instant. The symbol $s_{k}[n]$, after scaling with power control coefficient $\sqrt{p_k}$, is fed to the low-resolution DAC. Its distorted output, based on the Bussgang model~\cite{demir_bussgang_21}, is given as $    s_{\text{DAC},k}[n] = \alpha_{d,k}\sqrt{p_k}s_{k}[n]+\upsilon_{\text{DAC},k}[n]
$. Here $\rho_{d,k}$ is the DAC distortion factor and $\upsilon_{\text{DAC},k}[n]$ is the DAC quantization noise~\cite{demir_bussgang_21}. It is uncorrelated with the information signal $s_{k}[n]$, has a zero mean and variance $\rho_{d,k}\alpha_{d,k}p_{k}$. The DAC output $s_{\text{DAC},k}[n]$ is fed to the low-cost hardware-impaired RF chain, whose distorted output, based on the EVM model, is~\cite{mMIMObook2017}:\vspace{-3pt}
\begin{align} \setcounter{equation}{7}
    s_{\text{RF},k}[n]= s_{\text{DAC},k}[n] + \xi_{\text{RF},k}[n].  
\end{align}
 The term $\xi_{\text{RF},k}[n]$ is the transmit RF hardware impairment of the $k$th UE, and has pdf $\mathcal{CN}(0,\kappa_{t,k}^{2}(\mathbb{E}\{s_{\text{DAC},k}[n](s_{\text{DAC},k}[n])^{H}\}))$,  with $\kappa_{t,k}$ being the transmit EVM~\cite{Masoumi_CF_HW_20}. 
The $m$th AP receives the following sum signal at its antenna: $\yv_m[n] = \sum_{k=1}^{K}\hv_{mk}[n]s_{\text{RF},k}[n]$.  The AP feeds this receive signal to its hardware-impaired RF chains, which distorts it as\cite{Masoumi_CF_HW_20}:\vspace{-2pt}
\begin{align} \label{eq_y_m_RF_HW}
    \yv_{\text{RF},m}[n]=  \yv_{m}[n]  +\boldsymbol{\eta}_{\text{RF},m}[n]+{\zv_{m}[n]}.
\end{align}
Here $\boldsymbol{\eta}_{\text{RF},m}[n]$, with pdf $\mathcal{CN}(\mathbf{0},\kappa_{r,ap}^{2}\Wmat^{m}[n])$, is the receiver hardware distortion, and $\Wmat^{m}[n] =\text{diag}(\mathbb{E}\{\yv_{m}[n](\yv_{m}[n])^{H}| \hv_{mk}[n]\})$. The term $\zv_{m}[n]$ with pdf $ \mathcal{CN}(\boldsymbol{0},\mathbf{I}_N)$ is the $a$th AP AWGN. The RF chain output is then fed to the dynamic-resolution ADCs, whose noisy output, based on the Bussgang model~\cite{demir_bussgang_21}, is given as:\vspace{-3pt}
\begin{equation}
 \yv_{\text{ADC},m}[n]= \Amat_{m}\yv_{\text{RF},m}[n] + \nv_{\text{ADC},m}[n].
\end{equation}
The matrix $\Amat_m \!=\! \text{diag}\{1-\rho_{a,1}^m,\cdots,1-\rho^m_{m,N}\}$, with $\rho^m_{a,i}$ being the ADC distortion factor for the $i$th antenna. The vector $\nv_{\text{ADC},m}$ is the quantization noise which is uncorrelated with the input signal $\yv_{\text{RF},m}[n]$. It has a zero mean and covariance $\boldsymbol{\Theta}_m = \Bmat_{m}\Smat_{m}[n]$ where $\Bmat_{m}\!=\Amat_{m} (\Imat_N \!-\! \Amat_{m})$ and $\Smat_{m}[n]=\text{diag}(\E\{{\yv_{\text{RF},m}[n]}({\yv_{\text{RF},m}[n]})^{H}|\hv_{mk}[n]\})$~\cite{demir_bussgang_21}.\newline 
\textbf{Two-layer decoding:} The CF system considered herein employs two-layer decoding to mitigate the IUI. In the first-stage, each AP combines its received signal by using the local channel estimates, which mitigates a part of the IUI. To mitigate the residual IUI,  all APs send their locally-combined received signal to the CPU, which  performs the second-layer LSFD. The CPU computes the LSFD weights based on the large-scale fading coefficients, which it can easily compute at they remain constant for $100$s of coherence intervals \cite{Ozlem_CFwpt_21}. 
The $m$th AP first uses channel estimate $\Hat{\hv}_{mk}[\lambda]$ to combine the distorted received signal as 
$\Breve{s}_{km}[n] = \Hat{\hv}^{H}_{mk}[\lambda]{\yv_{\text{ADC},m}[n]}$.
It sends its combined signal $\Breve{s}_{km}[n]$ to the CPU which performs second-layer LSFD as  $\Hat{s}_{k}[n]\!= \!\sum_{m=1}^{M}a^{*}_{mk}[n]\Breve{s}_{km}[n]$. Here $a_{mk}$ is the complex~LSFD coefficient of the $m$th AP and the $k$th UE link. This reduces the IUI by weighing the received signals from all APs. We now re-express the signal $\Hat{s}_{k}[n]$ as given in \eqref{eq_signal_at_CPU} at
the top the page and use this to calculate the closed-form SE.

\vspace{-0.2cm}
\section{Spectral efficiency Analysis} 
We derive the closed-form SE expression by using the \textit{use-and-then-forget} UaTF technique~\cite{Zheng_aging_correlated_21,mMIMObook2017}. It expresses the LSFD-combined  signal at the AP $\Hat{s}_{k}[n]$ in~\eqref{eq_signal_at_CPU}, as  follows:\vspace{-3pt}
\begin{align}
 \Hat{s}_{k}[n]= \! \sum\limits_{m=1}^{M} \!\bar{a}_{mk}^{\rho *}[n]\sqrt{p_{k}}\E\Big\{\hat{\hv}_{mk}^{H}[\lambda]\Amat_{m}\hv_{mk}[\lambda] \Big\} s_k[n]   + \varpi_{k,n}. \notag
\end{align}\vspace{-1pt}
The effective noise term $\varpi_{k,n}$ consists of  all the terms given in~\eqref{eq_signal_at_CPU}, 
except the first term and the beamforming uncertainty term $\overline{\text{BU}}_{k,n} \!=\!\sum_{m=1}^{M}\bar{a}_{mk}^{\rho *}[n]\sqrt{p_{k}}\big(\hat{\hv}_{mk}^{H}[\lambda]\Amat_{m}\hv_{mk}[\lambda]\! - \, \E\big\{\!\hat{\hv}_{mk}^{H}[\lambda]\Amat_{m}\hv_{mk}[\lambda]\! \big\} \big)$.
The first term in $\Hat{s}_{k}[n]$, which is used for signal detection,  requires only long channel information~\cite{Zheng_aging_correlated_21}. Using central limit theorem, the effective noise can be approximated as  worst-case Gaussian~\cite{mMIMObook2017}. 
A lower bound to calculate the SE of a hardware-impaired CF mMIMO system with spatially-correlated Rician fading and channel aging for a given LSFD weights as  \cite{mMIMObook2017}: $\text{SE}_{k,n}\!= \text{log}_{2}(1+ \overline{\text{SINR}}_{k,n})$, where\vspace{-3pt} 
\begin{align} \setcounter{equation}{11}
\overline{\text{SINR}}_{k,n} \!=\!  \frac{\overline{\text{DS}}_{k,n}}{\begin{Bmatrix} \overline{\text{BU}}_{k,n}\!+\overline{\text{CA}}_{k,n} \!+\! \sum\limits_{i\neq k}^{K}\!\overline{\text{IUI}}_{ki,n} \!+ \overline{\text{DAC}}_{k,n} \\ \!+\overline{\text{TRF}}_{k,n}  \!+\overline{\text{RRF}}_{k,n} \!+\overline{\text{ADC}}_{k,n} \!+\overline{\text{NS}}_{k,n}\end{Bmatrix}\!\!}. \!\!  \label{eq_SE_expectations}
\end{align}
\begin{figure*}[h]
	\centering\vspace{+0.02in}
	\begin{subfigure}{.24\textwidth}
		\centering
		\includegraphics[width=\linewidth,height=\linewidth]{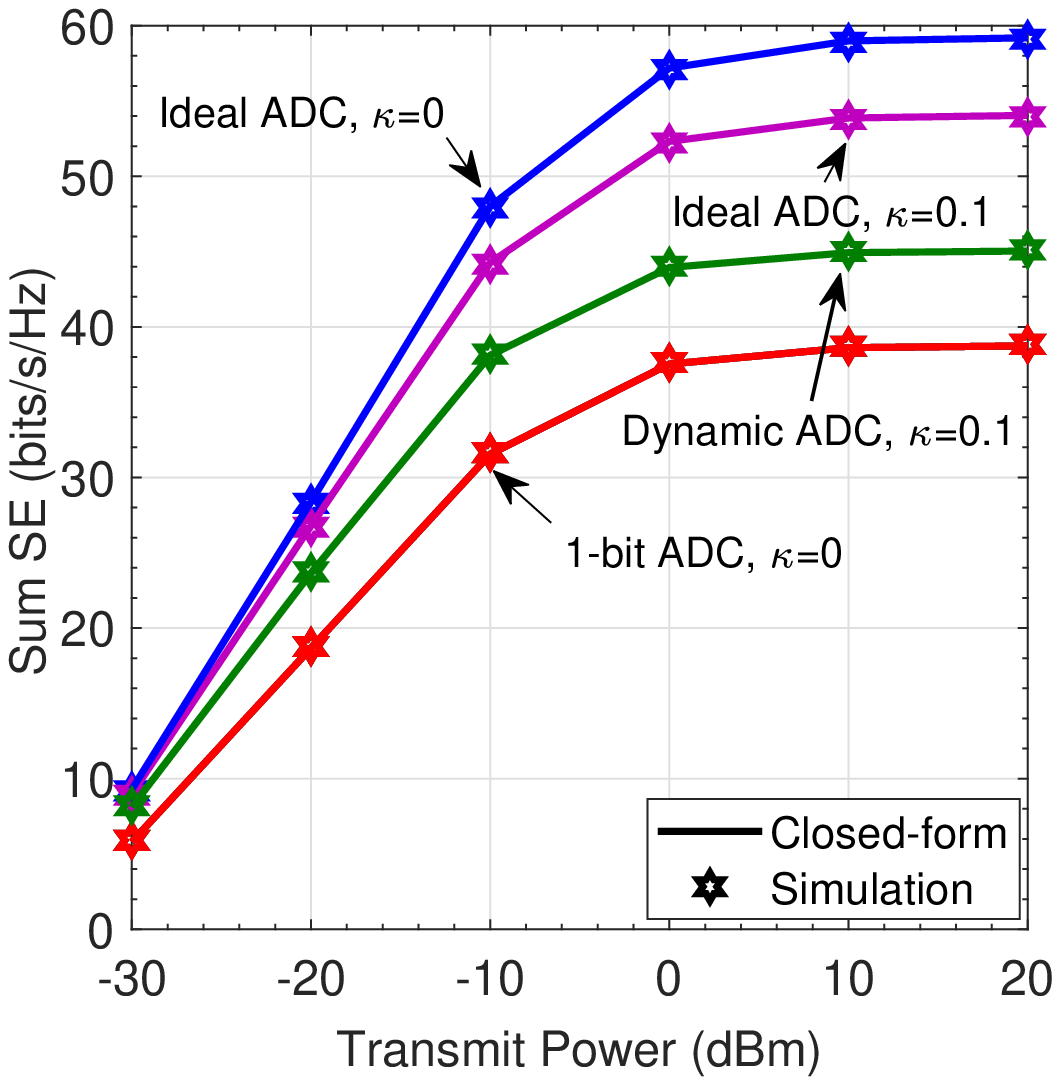}\vspace{-7pt}
		\caption{ \small} 
		\label{fig:SE_validation_ideal_dynamic_ADC}
	\end{subfigure}
	\begin{subfigure}{.24\textwidth}
		\centering
		\includegraphics[width=\linewidth,height=\linewidth]{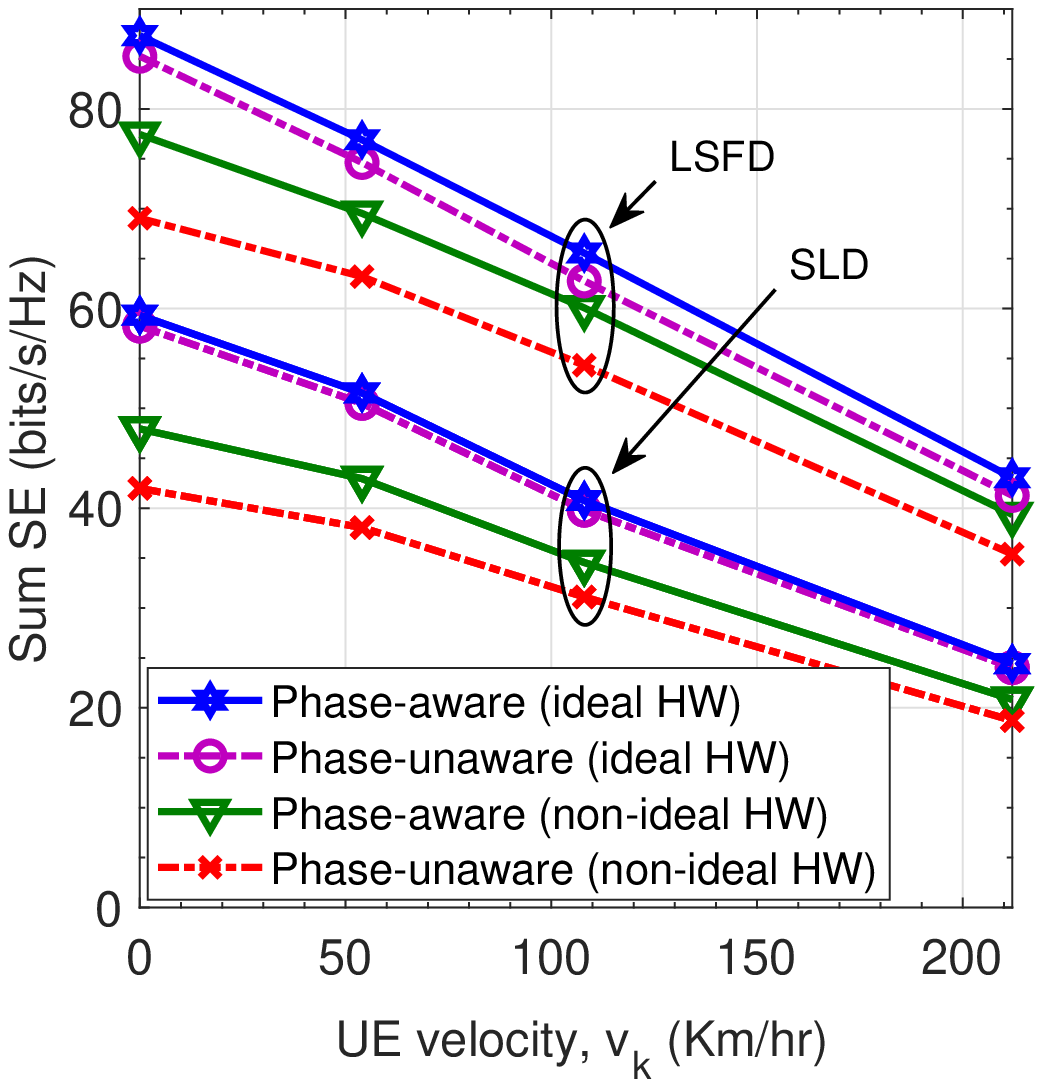}\vspace{-7pt}
		\caption{\small }
		\label{fig_SE_vs_Velocity_phase_aware_unaware} 
	\end{subfigure}
	\begin{subfigure}{.24\textwidth}
		\centering
		\includegraphics[width=\linewidth,height=\linewidth]{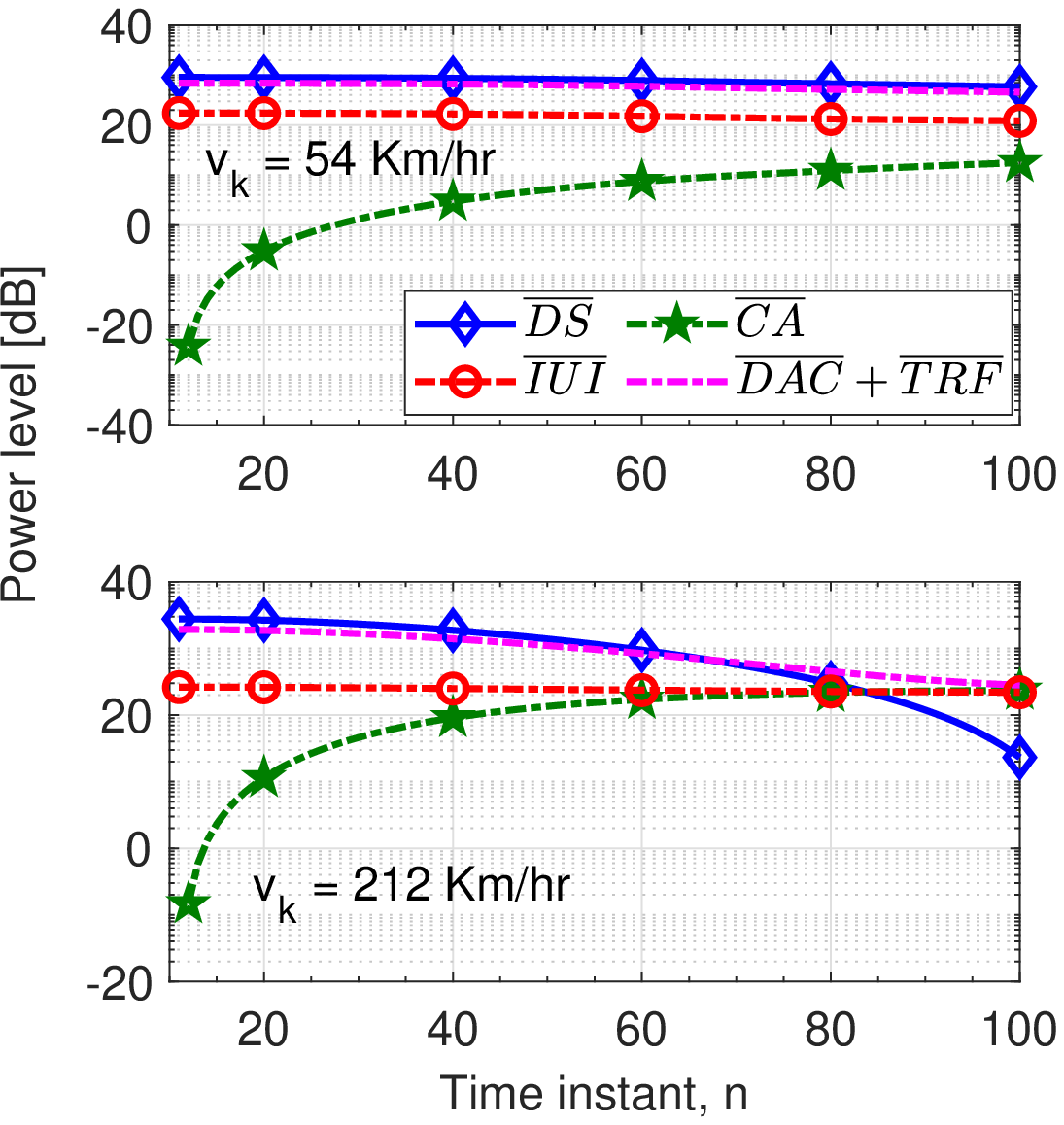}\vspace{-7pt}
		\caption{\small }
		\label{fig_CDF_LSFD_each_term_power} 
	\end{subfigure}
	\begin{subfigure}{.24\textwidth}
		\centering
		\includegraphics[width=\linewidth,height=\linewidth]{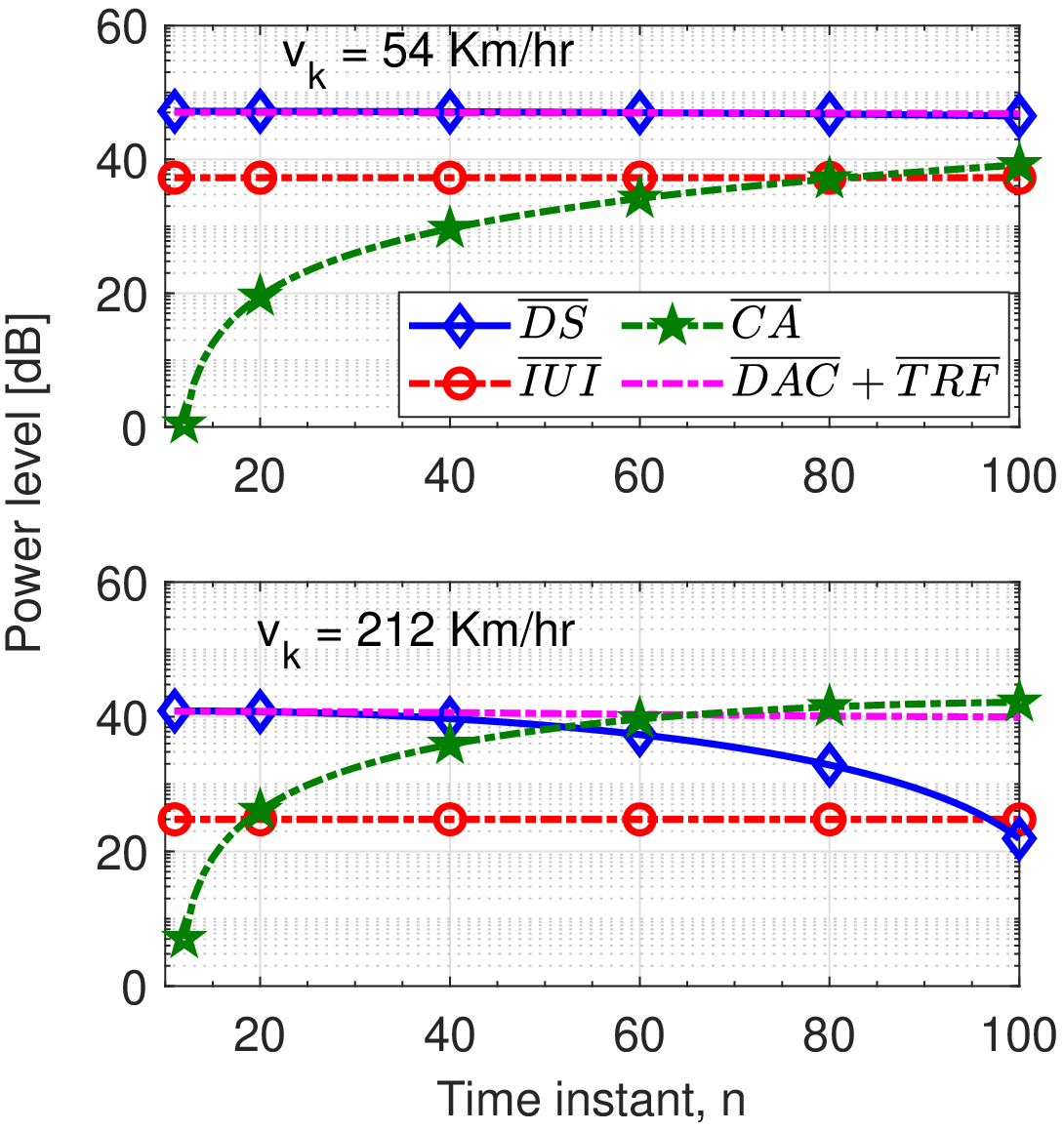}\vspace{-7pt}
		\caption{\small }
		\label{fig_CDF_SLD_each_term_power}
	\end{subfigure}
	\vspace{-6pt}
	\caption{a) Validation of closed-form SE; b) SE with phase aware and unaware estimators;  Effect of channel aging on signal terms for c) LSFD; and d) SLD. \vspace{-10pt}} 
	\label{fig:test}	
\end{figure*}
The term  $\overline{\text{DS}}_{k,n}$ is the desired signal strength, $\overline{\text{BU}}_{k,n}$ is the beamforming uncertainty strength, $\overline{\text{IUI}}_{ki,n}$ is the interference strength,  $\text{DAC}_{k,n}$ is the UE DAC impairment strength,  $\overline{\text{TRF}}_{k,n}$ is the UE RF impairment strength,  $\overline{\text{RRF}}_{k,n}$ is the AP RF impairment strength,  $\overline{\text{ADC}}_{k,n}$ is the AP ADC impairment noise and $\overline{\text{NS}}_{k,n}$ is the AWGN strength. These terms are defined~as in Table~\ref{table_DS_IUI_HW_ADC}.
Here , 
$ \bar{a}_{mk}^{\bar{\rho} *}[n] = a_{mk}^{*}[n]\alpha_{d,k}\rho_{k}[n_{\lambda}]$.
These expectations next need to be computed to derive the closed-form SE expression.
The presence of i) channel aging; ii) spatially correlated Rician channel with phase-shifts; iii) RF impairments and dynamic ADC architecture at the APs; iv) RF impairments and low-resolution DAC at the UEs; and v) LSFD, significantly complicates their computation, when compared with existing CF channel aging literature~\cite{Zheng_aging_correlated_21}, which considered Rayleigh fading with ideal hardware. Their simplification requires novel mathematical results, and are not a straightforward extension of ~\cite{Zheng_aging_correlated_21}. We next derive the lower bound in the following theorem.  
\begin{theorem} \label{theorem_lower_bound}
The closed-form SE expression, for an arbitrary number of antennas with channel aging,  RF and ADC/DAC impairments both at the APs and the UEs, LSFD and spatially-correlated Rician channel with phase-shifts,  is given as\vspace{-2pt} 
\begin{align}\label{eq_SINR_k}
\text{SE}_{sum}=\frac{1}{\tau_c}\sum\limits_{n=\lambda}^{\tau_c}\sum\limits_{k=1}^K\log_2\Big(1+   \frac{\alpha_{d,k}^{2}p_{k}|\av_{k}^{H}[n]\boldsymbol{\delta}_{kn}|^{2}}{ 
    \av_{k}^{H}[n] \boldsymbol{\Delta}_k \av_{k}[n]}\Big). 
\end{align}
Here $\boldsymbol{\Delta}_k \!=\! \Big(\!\sum\limits_{i=1}^{K} \!\big( \boldsymbol{\Xi}_{kin} \!+\!  \big(\rho_{d,i} \!+\!  \kappa_{t,i}^{2} \big)\alpha_{d,i}p_{i}\Cmat_{kin}  \! + \Dmat_{kin}\! \big) \!+ \!\boldsymbol{\Lambda}_{kn}$ \\
 $+  \sum\limits_{i\neq k}^{K} \alpha_{d,i}^{2}p_{i}\Cmat_{kin} + \Bmat_{kn}    
 + \text{diag}\big(\sigma^2\text{tr}\big(\bar{\boldsymbol{\Gamma}}_{mk}\Bmat_{m} \big) \big) +  \sigma^2\Qmat_{k} \Big)$.
with $\boldsymbol{\delta}_{kn}=[\delta_{kn}^{1},\hdots,\delta_{kn}^{M}]^{\text{T}}$, 
$\Bmat_{kn}=\text{diag}([\bv_{k,n}^{1},\hdots,\bv_{k,n}^{M}])$, $\Dmat_{kin} = \text{diag}(d^{1}_{kin},\cdots, d^{M}_{kin})$. The matrix $\Cmat_{kin}$ diagonal and off-diagonal elements are given as $c_{kin,m}$ and $c_{kin,mm^{'}}$ respectively. 
The terms $\delta_{kn}^{m}$, $\bv_{kn}^{m}$, $\boldsymbol{\Lambda}_{kn},\Cmat_{kin},\boldsymbol{\Xi}_{kin},\Dmat_{kin},\Qmat_{k} $ are given in Appendix~\ref{Appendix_theorem}. 
\end{theorem}
{We see from \eqref{eq_SINR_k} that the SINR expression is a generalized Rayleigh quotient with respect to $\av_{k}[n]$. 
The optimal LSFD coefficients are given~as
$    \av_{k}[n]\!= \! \boldsymbol{\Delta}_k ^{\!-1} \boldsymbol{\delta}_{kn}$~\cite[Lemma~B.10]{mMIMObook2017}. 
\begin{corollary} {We can simplify the sum-SE expression in \eqref{eq_SINR_k} for ideal hardware, and show that it matches with the existing work~\cite{HQNgo2017,Zheng_aging_correlated_21}. This not only validates our results theoretically but also shows that the current work subsumes the existing work~\cite{HQNgo2017,Zheng_aging_correlated_21}.} By setting i) Rician factor $K_{mk}\rightarrow 0$; ii) $\Rmat_{mk} = \beta_{mk}\Imat_{N}$; and iii) ideal RF hardware and high ADC/DAC resolution, the $\overline{\text{SINR}}_{kn}$ expression is\vspace{-3pt} 
\begin{align}\label{eq_SINR_k_rayl_ideal_HW}
   & \overline{\text{SINR}}_{k,n} \!=\\[-10pt]
    &\frac{p_{k}\rho_{k}^{2}[n_{\lambda}]|\sum\limits_{m=1}^{M}a^{*}_{mk}[n] N \bar{\gamma}_{mk}  |^{2}}
   { \begin{Bmatrix} \sum\limits_{i=1}^{K}\sum\limits_{m=1}^{M}\!\big|a_{mk}^{*}[n]\big|^{2}p_{i}N\bar{\gamma}_{mk}\beta_{mi} \! + \sigma^{2}\!\sum\limits_{m=1}^{M}\!|a_{mk}^{*}[n]|^{2}N\bar{\gamma}_{mk} \\[-5pt]
    + N^{2}\!\sum\limits_{i\in\mathcal{P}_{k}}\!p_{i}\rho_{i}^{2}[n_{\lambda}]\Big|\!\sum\limits_{m=1}^{M}\!a_{mk}^{*}[n]\sqrt{\bar{\gamma}_{mk}\bar{\gamma}_{mi}}\Big|^{2}  \end{Bmatrix} }. \nonumber
\end{align}
This above expression matches with~\cite[Eq. (21)]{Zheng_aging_correlated_21}. Also, with $\rho_k[n_{\lambda}] \!=\!1$, $N\!=\!1$, $\kappa\!=\!0$, $\rho_{d,i}\!=\!0$ and $a_{mk}[n] \!= 1/M$, the expression in \eqref{eq_SINR_k_rayl_ideal_HW} matches with~\cite[Eq. $(27)$]{HQNgo2017}. 
\end{corollary}

\begin{table*}[bp] 
  \centering \footnotesize
    \caption{Closed-form expressions obtained in Theorem~\ref{theorem_lower_bound}.}\vspace{-7pt}
   \label{appendix_Closed_form1}
  \begin{tabular}{|c|c|} 
 \hline
\makecell{For $i \in \mathcal{P}_{k}$, \\ $d_{kin}^{m}$}  &  \makecell{$ \kappa_{r,m}^{2}(1+\kappa_{t,i}^{2})\alpha_{d,i}p_{i} \Big( \sum_{j\in\mathcal{P}_{k}}(1+\kappa_{t,j}^{2})(1-\rho_{d,j})\tilde{p}_{j}\Big( \text{tr}\left(\left(\Bmat_{m} +\kappa_{r,m}^{2}\Amat_{m}\right)\text{diag}(\bar{\Rmat}_{mj}) \Pmat_{mk}\text{diag}\bar{\Rmat}_{mi})\Pmat_{mk}^{H}  \right)  $
\\ $+ \text{tr}\left(\bar{\Rmat}_{mj}\Amat_{m}\Pmat_{mk}  \text{diag}(\bar{\Rmat}_{mi})\Pmat_{mk}^{H}\Amat_{m}\right)\Big) + \sigma^{2}\text{tr}\left( \Amat_{m}\Pmat_{mk}  \text{diag}(\bar{\Rmat}_{mi})\Pmat_{mk}^{H}  \right) + (1+\kappa_{t,i}^{2})(\alpha_{d,i})\tilde{p}_{i}\rho_{i}^{2}[n-\lambda]\rho_{i}^{2}[\lambda-t_k]  $ 
\\ $\times \Big( \text{tr}\left(\left(\Bmat_{m}+\kappa_{r,m}^{2}\Amat_{m}\right)\text{diag}(\Rmat_{mi}\Pmat_{mk}^{H}) \Pmat_{mk}\Rmat_{mi} \right) + 2\text{real}\left\{ \text{tr}\left( \bar{\hv}_{mi}\bar{\hv}_{mi}^{H}\Amat_{m}\Pmat_{mk}\text{diag}\left( \Pmat_{mk}^{H}\Amat_{m}\Rmat_{mi} \right)      \right)\right\}$ 
\\ $ + \text{tr}\left( \Rmat_{mi}\Amat_{m}\Pmat_{mk}\text{diag}\left( \Pmat_{mk}^{H}\Amat_{m}\Rmat_{mi} \right)\right)+2\text{real}\left\{\text{tr}\left( (\Bmat_{m}+\kappa_{r,m}^{2}\Amat_{m})\left\{ (\bar{\hv}_{mi}\bar{\hv}_{mi}^{H})\odot\Pmat_{mk}\right\}\left\{ (\Rmat_{mi})\odot\Pmat_{mk}^{H}\right\}  \right)   \right\}  \Big) \Big)$. }  \\   
 \hline 
 \makecell{For $i \in \mathcal{P}_{k}$, \\ $c_{kin,m}$} & \makecell{$\sum_{j \in\mathcal{P}_{k}}\!(1\!+\! \kappa_{t,j}^{2})\alpha_{d,j}\tilde{p}_{j}\text{tr}\Big(\!\! \left(\Bmat_{m}\!+\!\kappa_{r,m}^{2}\Amat_{m} \right) \text{diag}\left(  \bar{\Rmat}_{mj}  \!\right) \Pmat_{\!mk} \bar{\Rmat}_{mi}  \Pmat_{\!mk}^{H}  \Big) \! +\! \text{tr}\left(\sigma^{2}\!\Amat_{m}\Pmat_{\!mk} \bar{\Rmat}_{mi}  \Pmat_{\!mk}^{H}  \right) +(1+\kappa_{t,i}^{2})\alpha_{d,i}\tilde{p}_{i}\rho_{i}^{2}[t_k^{\lambda}] $ 
 \\ $ \times \Big(2\text{real} \left\{ \text{tr}\left(\bar{\hv}_{mi}\bar{\hv}_{mi}^{H}\Pmat_{mk}^{H}  \left(\Bmat_{m}+\kappa_{r,m}^{2}\Amat_{m} \right) \text{diag}\left( \Pmat_{mk}\Rmat_{mi}\right) \right) \right\} + \text{tr}\left(\text{diag}\left(\Rmat_{mi}\Pmat_{mk}^{H} \right)(\Bmat_{m} +\kappa_{r,m}^{2}\Amat_{m}) \text{diag}\left(\Pmat_{mk}\Rmat_{mi} \right)  \right) \Big)$ 
 \\ $\sum_{j\in\mathcal{P}_{k}} (1-\rho_{d,j})(\rho_{d,j}+\kappa_{t,j}^{2})\tilde{p}_{j} \text{tr}\left( \bar{\Rmat}_{mj} \Amat_{m}\Pmat_{mk}\bar{\Rmat}_{mi} \Pmat_{mk}^{H}\Amat_{m}\right)  
    +\alpha_{d,i}(\rho_{d,i}+\kappa_{t,i}^{2})\tilde{p}_{i}\rho_{i}^{2}[t_k^{\lambda}] $ 
    \\ $\Big(  \left| \text{tr} \left( \Rmat_{mi}\Amat_{m}\Pmat_{mk}\right)    \right|^{2} 
     +2\text{real} \big( \bar{\hv}_{mi}^{H}\Amat_{m}\Pmat_{mk}\boldsymbol{{\Bar{h}}_{mi}} \text{tr} \left( \Rmat_{mi}\Pmat_{mk}^{H}\Amat_{m} \right)  \big) \Big) +\Bar{\rho_{i}}^{2}[n-\lambda]\text{tr}\left(\boldsymbol{\overline{\Gamma}}_{mk}\Amat_{m}\bar{\Rmat}_{mi}\Amat_{m}  \right)$} \\
 \hline
 \multicolumn{2}{|c|}{For $ \vphantom{\sum\limits_{j\in}^{k}} i \in \mathcal{P}_{k}$, $c_{kin,mm^{'}} = \rho_{i}^{2}[n_{\lambda}] \alpha_{d,i}^{4}\rho_{i}^{2}[t^{\lambda}_k]\tilde{p}_{i}^{2}\text{tr}\left(\bar{\Rmat}_{mi}\Amat_{m}\boldsymbol{\Psi}_{mk}\Amat_{m}\bar{\Rmat}_{mk}\Amat_{m} \right) \text{tr}\big(\bar{\Rmat}_{m^{'}i}\Amat_{m^{'}}\boldsymbol{\Psi}_{m^{'}k}\Amat_{m^{'}}\bar{\Rmat}_{m^{'}k}\Amat_{m^{'}} \big)$; For $i \notin \mathcal{P}_{k}$, $c_{kin,mm^{'}} =0$.}\\[-4pt] 
  \hline
  \multicolumn{2}{|c|}{$\vphantom{\sum\limits_{j\in}^{k}}\boldsymbol{\delta}_{kn}^m \!= \!\rho_{k}[n_{\lambda}]\text{tr}\left(\Amat_{m} \boldsymbol{\bar{\Gamma}}_{mk} \right)$; For $i \notin \mathcal{P}_{k}$, $d_{kin}^{m} \!=\! \kappa_{r,m}^{2}(1+\kappa_{t,i}^{2})\alpha_{d,i}p_{i} \text{tr}\big( \boldsymbol{\overline{\Gamma}}_{mk}\Amat_{m}\text{diag}\big({\bar{\Rmat}_{mi}} \big)\Amat_{m} \big)$, $c_{kin,m}\!=\!\text{tr}\big(\boldsymbol{\overline{\Gamma}}_{mk}\Amat_{m}\bar{\Rmat}_{mi}\Amat_{m}\big)$. } \\[-6pt] 
  \hline
 \multicolumn{2}{|c|}{ $\vphantom{\sum\limits_{j\in}^{k}}\boldsymbol{\Lambda}_{kn}\! = \!\text{diag}\big( \bar{\rho}_{k}^{2}[n_{\lambda}]\alpha_{d,k}^{2}p_k\text{tr}\big( \boldsymbol{\bar{\Gamma}}_{mk}\Amat_{m} \bar{\Rmat}_{mk}  \Amat_{m} \big) \big)$,  $\Qmat_{k} \!= \text{diag} \big( \text{tr}(\bar{\boldsymbol{\Gamma}}_{1k}\Amat_{1}^2) ,\cdots \!, \text{tr}(\bar{\boldsymbol{\Gamma}}_{Mk}\Amat_{M}^2) \big)$, $\Bmat_{k,n}\! = \text{diag}\big(\alpha_{d,k}^{2}p_{k}\big(c_{kkn,m} \!- | \text{tr}\big(\Amat_{m} \boldsymbol{\bar{\Gamma}}_{mk} \big)  |^{2}  \big) \big)$.} \\
  \hline
  
  \end{tabular}
\end{table*}
\section{Simulation Results}
We now numerically validate the derived closed-from SE in \eqref{eq_SINR_k} and investigate the i) effect of channel aging; ii) RF and ADC/DAC impairments; iii) Rician phase-shifts; and iv) LSFD. We consider a simulation setup where $M$ APs and $K$ UEs are randomly and uniformly distributed within an area of $1 \times 1 \; \text{km}^2$~\cite{HQNgo2017}. 
 We use the local-scattering model from~\cite{mMIMObook2017} to model the correlation matrix $\Rmat_{mk}$. We set  the noise variances $\sigma^2_m=\sigma^2_k =-94$~dBm, $M=64$~APs, $K\!=\!20$ UEs, $N\!=\!4$ antennas per AP, $v_k \!= \!54$ Km/hr and the pilot power $\tilde{p}_{k}=10$~dBm~\cite{mMIMObook2017}. We model the Rician factor and the large-scale fading coefficients as in~\cite[Eq. (84)]{Emil2019_rician_phase}.\newline
\textbf{\underline{Validation of closed-form SE:}} 
We first validate in Fig.~\ref{fig:SE_validation_ideal_dynamic_ADC} the derived closed-form SE expression in \eqref{eq_SINR_k} by comparing it with its simulated counterpart in~\eqref{eq_SE_expectations}. 
 For this study, we consider following RF impairment and ADC/DAC combinations: i) ideal RF with $\kappa \!=\!0$, ADC/DAC; ii) $\kappa_t\!=\!0.1$, ideal ADC/DAC; iii) $\kappa_t\!=\!0.1$, dynamic ADC resolution; iv) ideal RF, $1$-bit ADC/DAC resolution. 
 We see that for different combinations of RF and ADC/DAC impairments, the derived closed-form SE exactly matches with its numerical counterpart. This validates the derived closed form SE expression, which can thus be used for a realistic evaluation of hardware-impaired CF channel aging system.\newline
\textbf{\underline{Phase-aware and phase-unaware estimator performance:}} 
We now study in Fig.~\ref{fig_SE_vs_Velocity_phase_aware_unaware} the effects of phase-aware and phase-unaware MMSE channel estimator on the SE with two-layer LSFD and single-layer decoding (SLD) by varying the UE velocity $v_k$. The SLD can be obtained by considering $\av_{k}[n] = [1/M, \cdots, 1/M ]^T$ in \eqref{eq_SINR_k}.  
We first see that LSFD provide much higher SE than SLD. We also see that the phase-aware estimator has a higher SE than the phase-unaware estimator. This is because the phase knowledge helps in suppressing the interference~\cite{Emil2019_rician_phase}. Further, the SE gap between the two estimators reduces with increase in velocity.  The, consequent, faster channel aging increases the correlation factor $\bar{\rho}_k[n_{\lambda}]$, which increases the interference due to channel aging $\overline{\mbox{CA}}_{kn}$ in \eqref{eq_SE_expectations}. This mitigates the gain of phase-aware channel estimator.  {We also see that for both LSFD and SLD, the SE gap between the phase-aware and phase-unaware estimator increases with non-ideal hardware. 

\hspace{-10pt}{\textbf{\underline{Effect of channel aging:}}} 
{We now numerically study in Fig.~\ref{fig_CDF_LSFD_each_term_power} and Fig.~\ref{fig_CDF_SLD_each_term_power} the behavior of desired signal  power, channel aging power, IUI power, and UE RF and DAC distortion  power (labelled as $\overline{\text{DS}}$, $\overline{\text{CA}}$, $\overline{\text{IUI}}$ and $\overline{\text{DAC}}+ \overline{\text{TRF}}$, respectively), for LSFD and SLD schemes. We assume that $K/2$ UEs have a velocity of $v_k=54$~Km/hr, and the remaining $K/2$ UEs have $v_k=212$~Km/hr. From Fig.~\ref{fig_CDF_LSFD_each_term_power}, we observe the following:}
	\begin{itemize}[leftmargin = *]
 		\item {For high-velocity UEs, the $\overline{\text{DS}}$ and $\overline{\text{DAC}}+\overline{\text{TRF}}$ powers reduce with time (see the bottom subplot). This is because both these powers are functions of the temporal correlation coefficient $\rho_{k}[n_{\lambda}]$, whose value, for a high UE velocity,  reduces greatly with time. For low-velocity UEs, as shown in the top subplot, these powers are almost time-invariant.} 
		\item {For high-velocity UEs and $n \ge 60$, we see that $\overline{\text{DS}}$ reduces monotonically,  while $\overline{\text{DAC}}+\overline{\text{TRF}}$ power first tapers out, and then floors to a constant value. This is because the $\overline{\text{DS}}$ power decreases as $\rho_k^2[n_{\lambda}]$, which decreases monotonically with $n$. The transmit RF + DAC impairments reduce as $\epsilon + \rho_k^2[n_{\lambda}]$, with $\epsilon > 1$ being a constant. For  $n\geq 60$, the constant $\epsilon $ term dominates the reduction due to $\rho_k^2[n_{\lambda}]$, which makes the  $\overline{\text{DAC}}+\overline{\text{TRF}}$ impairment power floor to a constant value.} 
	\end{itemize}
We next compare Fig.~\ref{fig_CDF_LSFD_each_term_power} with Fig.~\ref{fig_CDF_SLD_each_term_power}, which plots the above power values for SLD. We see that for low-velocity UEs, LSFD has  a much lower $\overline{\text{IUI}}$, $\overline{\text{CA}}$  and $\overline{\text{DAC}}+\overline{\text{TRF}}$ power than SLD. This is because the channels of low-velocity UEs do not significantly age,  and consequently their channel estimates quality do not deteriorate. The LSFD can thus better suppress the IUI.  For high-velocity UEs, LSFD yields much lower $\overline{\text{CA}}$ values  than SLD, while both LSFD and SLD yield similar $\overline{\text{IUI}}$ values. This implies that LSFD can mitigate the effect of channel aging for high-velocity UEs, but not IUI. 
This study shows that LSFD can mitigate i) the effect of channel aging for low/high UE velocities;  and ii) IUI for low-velocity UEs.
\vspace{-0.1cm}
\section{Conclusion} \vspace{-0.1cm}
We derived and validated the closed-form SE expression for a hardware-impaired spatially-correlated Rician-faded CF mMIMO system with channel aging and LSFD. We numerically showed that the SE gap between the phase-aware and phase-unaware estimators reduces with increased UE velocity, and increases with hardware impairments. {We also showed that the LSFD effectively mitigates the interference due to channel aging for both low- and high-velocity UEs. It, however, mitigates inter-user interference only for the low-velocity UEs.}
\vspace{-0.2cm}
\appendices
\section{} \label{Appendix_theorem} \vspace{-0.2cm}
We skip	the derivation of the terms in the Theorem~\ref{theorem_lower_bound} due to page constraint and provide it in the technical report~\cite{CF_AGING_APPENDIX_DOC_22}. Their closed-form expressions are, however, given in Table~\ref{appendix_Closed_form1}.
\vspace{-5pt}
\bibliographystyle{IEEEtran}
\bibliography{IEEEabrv,CF_aging_ref}
\end{document}